\documentclass[twocolumn, showpacs, preprintnumbers, amsmath, amssymb] {revtex4-1}
\usepackage{times}
\usepackage{amsmath}
\usepackage{amssymb}
\usepackage{graphicx}
\usepackage{natbib}
\usepackage{psfrag}
 \usepackage{setspace}
 \usepackage{color}

\newcommand{\Da}{\mathrm{Da}}
\newcommand{\St}{\mathcal{S}}

\newcommand{\Rm}{\mathrm{R}_m}
\newcommand{\Rc}{\mathrm{R}_c}

\newcommand{\Cc}{\mathcal{C}}

\newcommand{\partiald}[2]{\frac{\partial #1}{\partial #2}} 
\newcommand{\uvec}{\mathbf{u}} 
\newcommand{\xvec}{\mathbf{x}} 
\newcommand{\nvec}{\mathbf{n}} 

\begin{document}

\title {Optimal and hysteretic fluxes in alloy solidification: Variational principles and chimney spacing}

\author{ A. J. Wells$^{1,4}$, J. S. Wettlaufer$^{1,2,4}$  \& S. A. Orszag$^{3,4}$ }

\affiliation{$^{1}$Department of Geology and Geophysics, Yale University, New Haven, CT, 06520, USA\\$^{2}$Department of Physics, Yale University, New Haven, CT, 06520-8109, USA\\$^{3}$Department of Mathematics, Yale University, New Haven, CT, 06520-8283, USA\\$^{4}$Program in Applied Mathematics, Yale University, New Haven, CT, 06520, USA}

\begin{abstract}
We take a numerical approach to analyze the mechanisms controlling the spacing of chimneys -- channels devoid of solid -- in two-dimensional mushy layers formed by solidifying a binary alloy.  Chimneys are the principal conduits through which buoyancy effects transport material out of the mushy layer and into the liquid from which it formed.  Experiments show a coarsening of chimney spacing and we pursue the hypothesis that this observation is a consequence of a variational principle: the chimney spacing adjusts to optimize material transport  and hence maximize the rate of removal of potential energy stored in the mushy layer.  The optimal solute flux increases approximately linearly with the mushy layer Rayleigh number.  However, for spacings below a critical value the chimneys collapse and solute fluxes cease, revealing a hysteresis between chimney convection and no flow.  
\end{abstract}

\maketitle

Variational principles constitute a cornerstone of physics because the trajectory of a system is determined from the extremum of the action.  A common example in classical physics is an action defined as the time integral of the Lagrangian.  However, variational principles for nonlinear dissipative systems constitute a topic of long standing debate because the non-conservation of phase space volume implies such systems are not Hamiltonian \cite[e.g.][]{Wisdom}.  
Successful examples include turbulent Rayleigh-B\'{e}nard convection, where a variational approach yields bounds on the heat flux that compare favorably with scaling arguments~\cite{Doering:2006tg}; a similar approach has been applied to shear driven turbulence~\cite{DOERING:1992vl}. In this Letter we consider how a variational principle can be applied to describe convection in a \emph{mushy layer}: a reactive porous medium formed during solidification of a binary alloy~\cite[][]{Worster:2000}. In addition to shedding light on the dynamics of nonlinear dissipative systems, this problem has direct applications in geophysical, geological and industrial settings. For example, mushy-layer convection is principally responsible for brine drainage from young sea ice and the consequent buoyancy forcing of the polar oceans~\cite[][]{Wettlaufer:1997rz}. 

Under common growth conditions morphological instability of the solid-liquid interface generates a mushy layer: a reactive porous medium of solid dendrites bathed in concentrated fluid. The interstitial fluid can become convectively unstable resulting in buoyancy-driven convection within the mushy layer
\cite{Worster:1997}.  Convection drives flow of solute depleted/enriched fluid into regions of high/low solute concentration, leading to local dissolution/growth of the solid matrix because fluid in the interstices adjusts to maintain local thermodynamic equilibrium. Regions of low solid fraction have high permeability, and hence flow focussing accelerates the growth of the instability. The nonlinear growth of this instability leads to the formation of channels of zero solid fraction, or \emph{chimneys}, which form the principal conduits for drainage of solute from the layer. Experiments show that under a constant solidification rate, the chimneys are regularly spaced~\cite[e.g.][]{Peppinetal:2008}, whereas during growth from a fixed temperature surface the mean spacing between chimneys increases over time as the mushy layer thickens~\cite{Wettlaufer:1997rz, Solomon:1998rc}.  

The onset of convection and local dissolution is predicted by linear and weakly nonlinear stability analyses~(reviewed in \cite{Worster:1997, Worster:2000} ), but after chimneys form a different theoretical approach is required to account for the combination of porous medium flow in the mushy region and pure liquid flow in the chimney. Previous analyses treated either an isolated chimney~\cite{Worster:1991} or modelled dynamics that arise with a periodic array of chimneys of imposed spacing~\cite{SchulzeWorster:1998, ChungWorster:2002}. These require the areal number density of chimneys to be specified {\em a-priori}, and thus any subsequent prediction of solute fluxes relies on an independent theoretical prediction of the spacing of chimneys. 

\begin{figure}
\psfrag{LIQUID}{\textsc{Liquid}}
\psfrag{MUSH}{\textsc{Mush}}
\psfrag{SOLID}{\textsc{Solid}}
\psfrag{CHIMNEY}{\textsc{Chimney}}
\psfrag{L}{$l$}
\psfrag{V}{$V$}
\psfrag{azt}{$2 a$}
\psfrag{zeq0}{$z=0$}
\psfrag{zeqhxt}{$z=h(x,t)$}
\psfrag{TCPhiPsi}{$T$, $C$, $\phi$, $\hat{\mathbf{u}}$}
\psfrag{CeqCETeqTE}{$C=C_E,\quad T=T_E$}
\psfrag{CeqCoTeqTo}{$C=C_0,\quad T=T_{\infty}$}
\includegraphics[height=3.5cm]{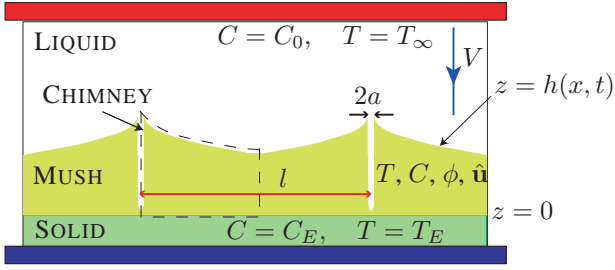}
\caption{(Color online) A two dimensional mushy layer formed between two heat exchangers pulled at velocity $V$ aligned with the gravitational acceleration $g\mathbf{k}$. The overlying liquid has constant far-field temperature $T_{\infty}$ and concentration $C_0$, and the mush solidifies at eutectic temperature $T_E$ and concentration $C_E$ at the lower boundary $z=0$. We assume a periodic array of chimneys, of dimensional width $2\hat{a}(z,t)$ and imposed spacing $l$, and exploit symmetry to solve for the properties of a mushy layer of thickness $h(x,t)$ within the dashed outline. The specific heat capacity $c_p$ and thermal diffusivity $\kappa$ are assumed constant across solid and liquid phases, and the fluid has dynamic viscosity $\mu$ and density $\rho_0 g \beta(C-C_E)$ for constant haline coefficient $\beta$ and reference density $\rho_0$. This occurs in aqueous $\mathrm{NH}_4 \mathrm{Cl}$ solidified from below and the dynamics and thermodynamics are ostensibly the same as ice forming above salt water.}
\label{fig:notation}
\end{figure}

We hypothesize that the chimney spacing adjusts to optimize drainage of potential energy from the mushy layer, and thus the system dynamics are determined by a variational principle that yields optimal solute fluxes. The resulting properties are determined numerically for two-dimensional steady-state solidification and we use this to reconcile behavior observed during transient growth. Fig.~\ref{fig:notation}  describes the two-component mixture of liquid concentration $C$ and temperature $T$ that is translated at a velocity $V$ between hot and cold heat exchangers.
Solidification depletes the liquid of solute, reducing the density and providing the buoyancy that drives convection. We investigate the behavior of a periodic array of chimneys within this system.  For a given chimney spacing $l$, we calculate the resulting solute flux.

We employ so-called ``ideal'' mushy layer theory which describes conservation of heat and 
solute, along with incompressible Darcy flow.  We draw upon an analysis of boundary conditions and their implications developed previously \cite{EmmsFowler:1994, SchulzeWorster:1998, ChungWorster:2002,  SchulzeWorster:2005}, and combine this with a fully time-dependent treatment and the hypothesized variational principle to reveal the new results presented here.
Within the mushy layer $T$ and $C$ are coupled by local thermodynamic equilibrium and hence lie on the liquidus curve $T=T_L(C)=T_E+\Gamma(C-C_E)$, where $\Gamma$ is constant, so that the local dimensionless temperature is $ \theta = \left[T-T_L(C_0)\right]/\Gamma \Delta C=\left(C-C_0\right)/\Delta  C$
where $C_0$ is the concentration in the liquid layer and $\Delta C=C_0-C_E$.  We solve for the dimensionless temperature $\theta$ and solid fraction $\phi$ and calculate the Darcy velocity $\mathbf{u}=(\psi_x,-\psi_z)$ by generating a vorticity equation for the dimensionless streamfunction $\psi$, assuming that the fluid density depends linearly on concentration and that the mushy layer has permeability $\Pi=\Pi_0(1-\phi)^3$.  Velocities, lengths and times are scaled by $V$,  $l_T=\kappa/V$ and $t_T=\kappa/V^2$ respectively from which we obtain six dimensionless parameters governing the system, 
\begin{align} \Rm&=\frac{\rho_0g\beta\Delta C\Pi_0 l_T}{\mu \kappa},  &  \Cc&=\frac{C_S-C_0}{\Delta C},& \Da &= \frac{\Pi_0 V^2}{\kappa^2},  \nonumber \\
 \theta_{\infty}&=\frac{T_{\infty}-T_L(C_{0})}{\Gamma \Delta C},& \St&=\frac{L}{c_p\Gamma \Delta C}, & \lambda&=\frac{V l}{2\kappa}. \label{eq:params2} \end{align}
The mushy layer Rayleigh number $\Rm$ describes the ratio of buoyancy to dissipation, and the Darcy number $\Da$ characterizes the mushy layer permeability. The Stefan number $\St$, concentration ratio $\Cc$ and scaled temperature $\theta_{\infty}$ characterize the imposed thermodynamic conditions.

Rather than solving directly for the overlying fluid layer, we apply a boundary layer approximation to describe its influence on the mushy layer. We assume constant pressure at the mush--liquid interface~\cite[][]{EmmsFowler:1994}, and that in the absence of solutal diffusion the fluid region has uniform concentration $C_0$ away from plumes exiting the mushy layer~\cite[][]{SchulzeWorster:1998}. The position of the mush--liquid interface is determined by the condition of marginal equilibrium $\theta=0$ at $z=h$, and hence continuity of salinity and normal heat fluxes give $\phi=0$ and $\left. \nvec \cdot \nabla T\right|_{-}^{+}=0$ at $z=h$. Applying a boundary layer approximation that balances advection and diffusion of heat across isotherms of curvature $\nabla \cdot \nvec$ yields
\begin{equation}\nvec \cdot \nabla \theta=\theta_{\infty} \left[  \nabla \cdot \nvec -\left(\uvec - \mathbf{k} \right) \cdot \nvec\right].  \label{eq:lidupdate} 
\end{equation}
The lower boundary $z=0$ is impermeable and fixed at the eutectic temperature ($\theta=-1$), and we apply symmetry conditions at the right hand boundary of the domain $x=\lambda$.

The boundary conditions at the chimney wall $x=a(z,t)$ play a key role in describing the flow. Chimneys are narrow $(a\ll1)$ so they can be represented by singular interface conditions at $x=0$. Lubrication theory applied to the flow in the narrow chimney yields the mass flux condition
\begin{equation}\psi=\left[\frac{a^3}{3\Da(1-\phi)^3}+a\right]\frac{\partial \psi}{\partial x} + \frac{3}{20}\frac{\Rm}{\Da} a^3 (\theta+1), \label{eq:chimneypsi} \end{equation}
where the pre-factor for the forcing has been calculated from a quadratic Polhausen approximation~\cite[][]{SchulzeWorster:1998, ChungWorster:2002}. Balancing the heat flux conducted into the chimney with that advected along the chimney yields
\begin{equation} \partiald{\theta}{x}=\psi \partiald{\theta}{z}. \label{eq:chimneyheatflux} \end{equation}
The chimney wall   is a free boundary with net outflow and radius $a(z,t)$ determined from the condition \cite[][]{SchulzeWorster:2005}
 \begin{equation} \partiald{\theta}{t}-\partiald{\theta}{z}+\uvec \cdot \nabla \theta=0. \label{eq:chimneymargeq} \end{equation}
 
 The system, including boundary conditions~\eqref{eq:lidupdate}--\eqref{eq:chimneymargeq}, was integrated numerically using second-order finite differences, with heat and concentration equations treated using semi-implicit Crank-Nicolson time-stepping. Elliptic equations for $\psi$ and $\theta$ were solved using multigrid iteration~\cite[][]{BriggsHensonMcCormick:2000, Adams:1989}. Finally, the chimney radius and mush-liquid interface position $h(x,t)$ were updated using relaxation.  The chimney radius was treated as a free boundary and updated at each spatial grid-point to reduce the error in~\eqref{eq:chimneymargeq}. The boundary layer approximation~\eqref{eq:lidupdate} leads to an unstable scheme for the corresponding free boundary problem for $h(x,t)$. Hence, we enforce a one parameter shape
 \begin{equation} h=h_1-\psi_c \left[ 1-\mathrm{cosh}\, \mu \left(\lambda-x\right)\right]/\left(\mu \mathrm{sinh}\, \mu \lambda\right),  \label{eq:hshape} \end{equation}
where $\psi_c$ is the streamfunction value at $\xvec=(0,h)$ \cite[e.g.,][]{SchulzeWorster:1998}.  To remove a temperature singularity at the chimney top this shape has a thermal boundary layer of width $1/\mu=\Rm^{-2/3}$. The parameter $h_1$ is adjusted to minimize the residual in satisfying~\eqref{eq:lidupdate} in a least squares sense, with $\theta=0$ enforced at $z=h$. Importantly, the time-dependent initial value problem was integrated to a steady state, for imposed values of the chimney half-spacing $\lambda=lV/2\kappa$ . The initial conditions were given either by a similarity solution with no fluid flow~\cite{Worster:1991} or by continuation from a previous steady state solution with different parameters. An arc length continuation scheme was also used to provide an alternative confirmation of the steady-state solution branches~\cite{Keller:1977}.
 
\begin{figure}
 \centering
    \psfrag{ls}{$\lambda_s$}
   \psfrag{lc}{$\lambda_{\cal O}$}
      \psfrag{lu}{$\lambda_u$}
    \psfrag{lambda}{$\lambda=l V/2 \kappa$}
   \psfrag{BF}{$F_S$}
    \psfrag{a}{\small (\textit{a})}
        \psfrag{b}{\small(\textit{b})}
         \psfrag{lambda2}{$\lambda$}
           \psfrag{lam}{\tiny$\lambda$}
            \psfrag{B2F}{\tiny$F_S$}
         \psfrag{Rm}{$\Rm$}
         \psfrag{R1}{$\mathrm{I}$}
         \psfrag{R2}{$\mathrm{II}$}
         \psfrag{R3}{$\mathrm{III}$}
         \psfrag{R4}{$\mathrm{IV}$}
                                                            \includegraphics[height=3.5cm]{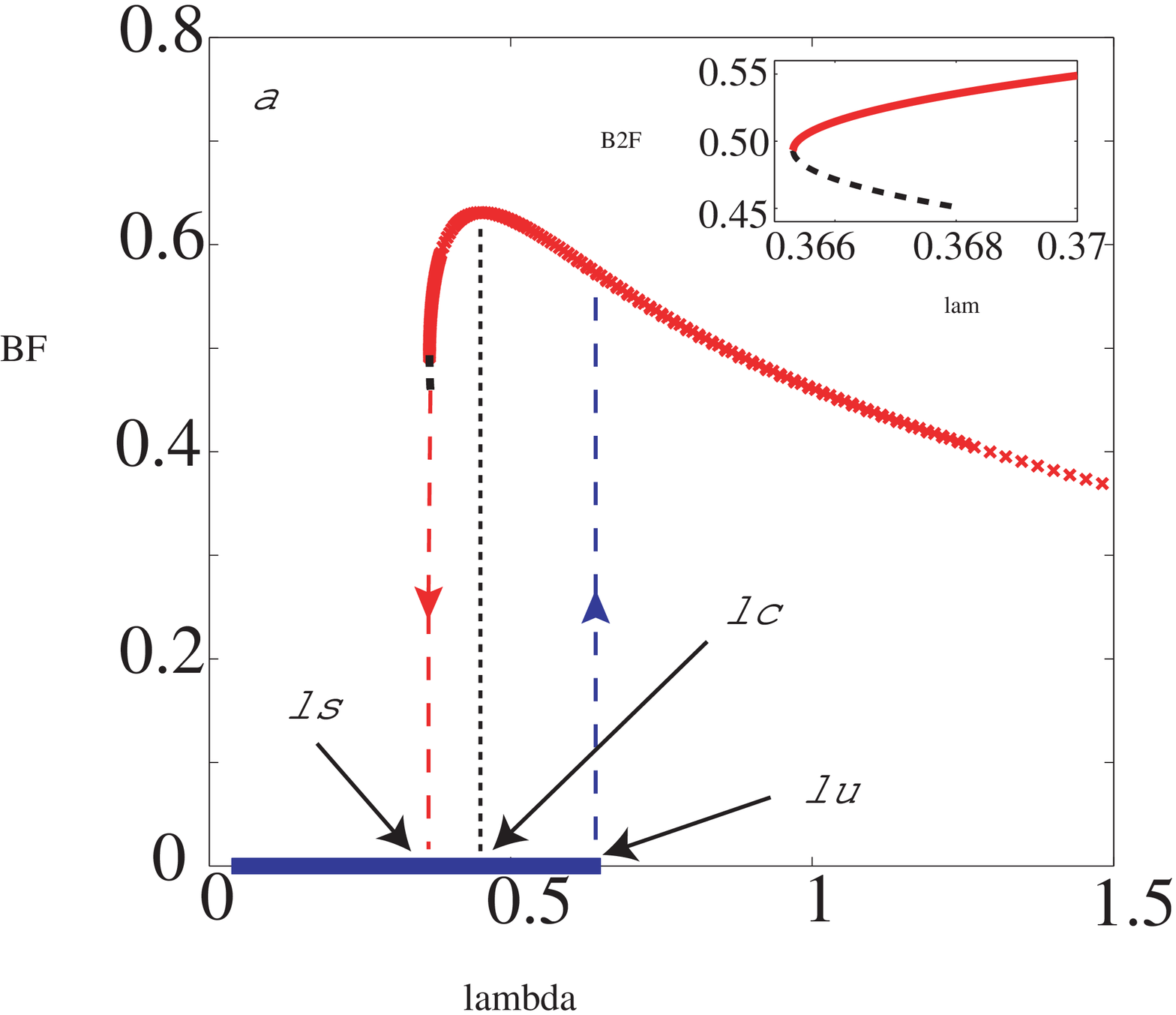}
                                                              \includegraphics[height=3.5cm]{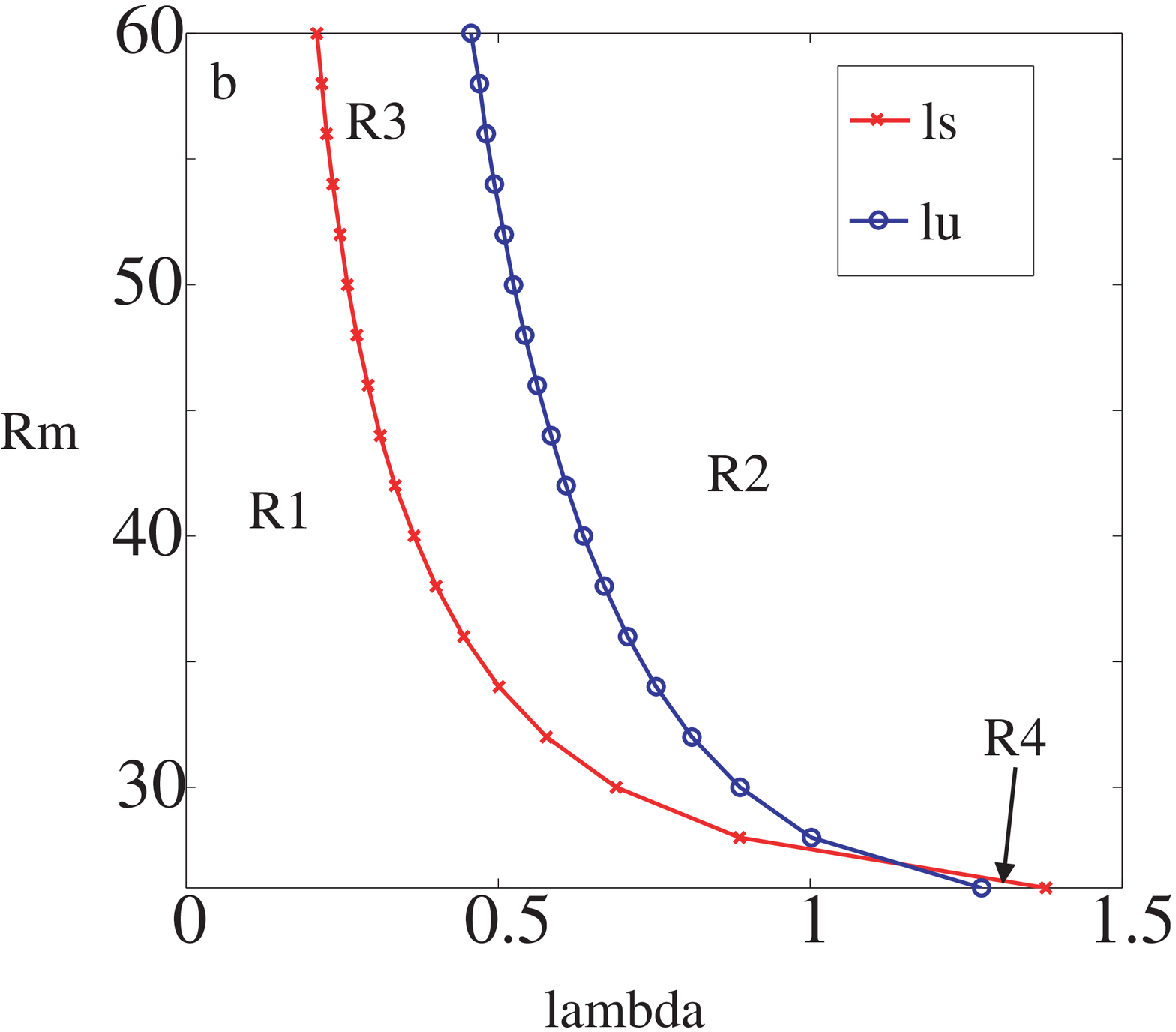}
\caption{(Color online) (a) Steady-state solute fluxes $F_S(\lambda)$ exiting a mushy layer vary with the chimney spacing $\lambda$. Two branches of solutions are determined and exhibit hysteresis; the upper branch results from convection with chimneys (red crosses) and the lower branch results from a state of no flow and hence $F_S(\lambda) = 0$ (blue squares). The upper branch exists for $\lambda>\lambda_s$ and the lower branch is unstable for $\lambda>\lambda_u$ with red and blue dashed lines indicating the solution trajectories at these points. A maximal solute flux ${F_S}_{\cal O}$ is attained for a chimney spacing $\lambda=\lambda_{\cal O}$, with weaker solute drainage at $\lambda>\lambda_{\cal O}$. The inset shows detail of the upper stable branch (red curve) and intermediate unstable branch (black dashed curve) in the vicinity of the stabilization point confirmed by arc length continuation. The calculations are for $\Rm=40$, $\Cc=15$, $\St=5$, $\theta_{\infty}=0.4$ and $\Da=5 \times 10^{-3}$ for consistency with previous studies~\cite[][]{ChungWorster:2002} which use properties for aqueous $\mathrm{NH}_4 \mathrm{Cl}$. The optimal solute flux is approximated by~(\ref{eq:BlowRmscale}-\ref{eq:BhiRmscale}), with $\gamma=0.03$ and $\Rc=20$ for these parameters. (b) Stability curves tracing the variation of $\lambda_s$ (red points) and $\lambda_u$ (blue points) with $\Rm$, with all other parameters held fixed. In  $\mathrm{I}$ only the lower branch is stable yielding no flow, and in $\mathrm{II}$ only the upper branch is stable yielding convection with chimneys. Hysteresis is observed with two steady states in $\mathrm{III}$. Both no flow and chimney convection states are unstable in $\mathrm{IV}$, and we observe a state of weak convection with no chimney.  }
\label{fig:brinefluxes}
\end{figure}

We investigate the influence of chimney spacing and convective strength on the mushy layer dynamics, using solutions for a range of $\lambda$ and $\Rm$ with $\Cc$, $\St$, $\theta_{\infty}$ and $\Da$ held fixed. First consider the variation of the chimney spacing wavelength $\lambda$ with the Rayleigh number held fixed at $\Rm=40$, noting that qualitatively similar behavior is observed for other values $27\lesssim \Rm \leq 60$.  The solute flux $F_S$ from the mushy layer is shown as a function of chimney spacing $\lambda$ in fig.~\ref{fig:brinefluxes}(a). 
There are two steady state branches,  a lower branch corresponding to a state of no flow, and an upper branch describing convection with chimneys. Depending on the choice of initial conditions, hysteresis  is found with one of two stable steady solutions  over a range $\lambda_s<\lambda<\lambda_u$.   A state of no flow remains stable for $\lambda<\lambda_u$, but becomes unstable for $\lambda>\lambda_u$ with the solution evolving in time to the upper branch of chimney convection. If we start on the upper branch and reduce $\lambda$ then chimney convection remains stable for $\lambda>\lambda_s$, but when $\lambda<\lambda_s$ chimneys collapse,  returning the system to a state of no flow. Fig.~\ref{fig:brinefluxes}(b) traces the stability boundaries $\lambda_s$ and $\lambda_u$ versus $\Rm$, and identifies regions of phase space with no flow ($\mathrm{I}$), chimney convection ($\mathrm{II}$) and both steady states ($\mathrm{III}$). Hence, starting in a state of chimney convection in region $\mathrm{III}$ and reducing $\lambda$ the system crosses the stability boundary $\lambda_s(\Rm)$, the flow is stabilized and chimney convection ceases. For $\Rm\lesssim27$, the stability curves cross and the nature of the solution changes. Additional calculations indicate a state of weak convection with no chimneys, and hence $F_S$=0, observed in region $\mathrm{IV}$, where both chimney convection and no flow states are unstable.

Because there are always sufficiently large wavelengths $\lambda$ available to trigger the instability of a state of no flow, we examine the the upper solution branch to find that  $F_S(\lambda)$ has a maximum at some critical wavelength $\lambda=\lambda_{\cal O}$.  Hence,  an optimal solute flux can be attained by varying the chimney spacing (Fig.~\ref{fig:brinefluxes}a). The solute flux weakens at large wavelengths $\lambda>\lambda_{\cal O}$, which can be understood by considering examples of the mushy layer properties at different chimney spacings.
 \begin{figure}
  \psfrag{xi}{$x$}
 \psfrag{x}{}
 \psfrag{z}{\hspace{-0.2cm} $z$}
 \psfrag{ph}{$\phi$}
 \psfrag{th}{$\theta$}
 \psfrag{eps}{$a$}
 \psfrag{a}{(\textit{a}) \; $\lambda>\lambda_{\cal O}$}
  \psfrag{b}{(\textit{b})}
     \psfrag{e}{(\textit{c}) \; $\lambda=\lambda_{\cal O}$}
      \psfrag{f}{(\textit{d})}
      \psfrag{c}{(\textit{c})}
      \psfrag{d}{(\textit{d})}
 \includegraphics[height=5.6cm]{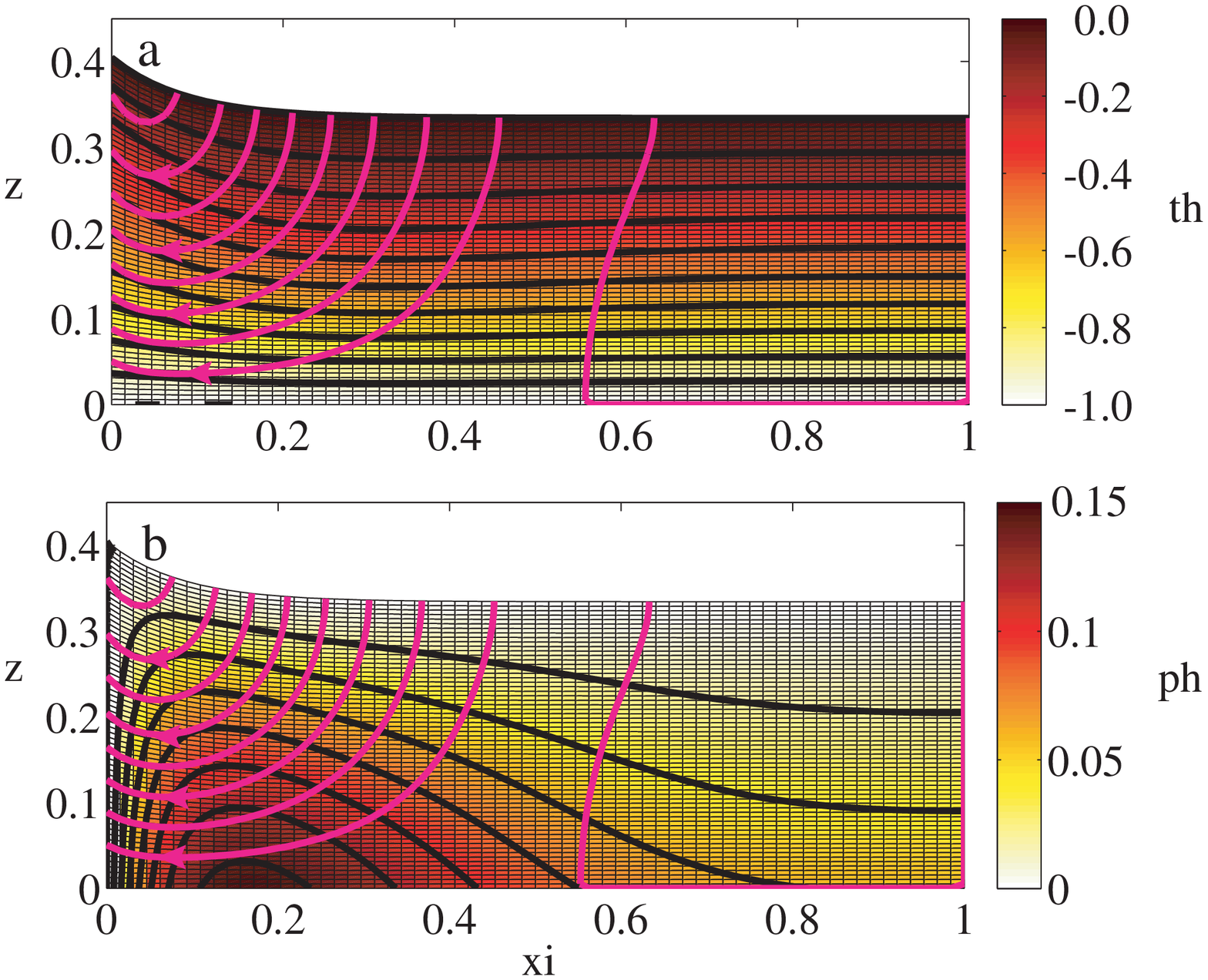} 
    \includegraphics[height=2.8cm]{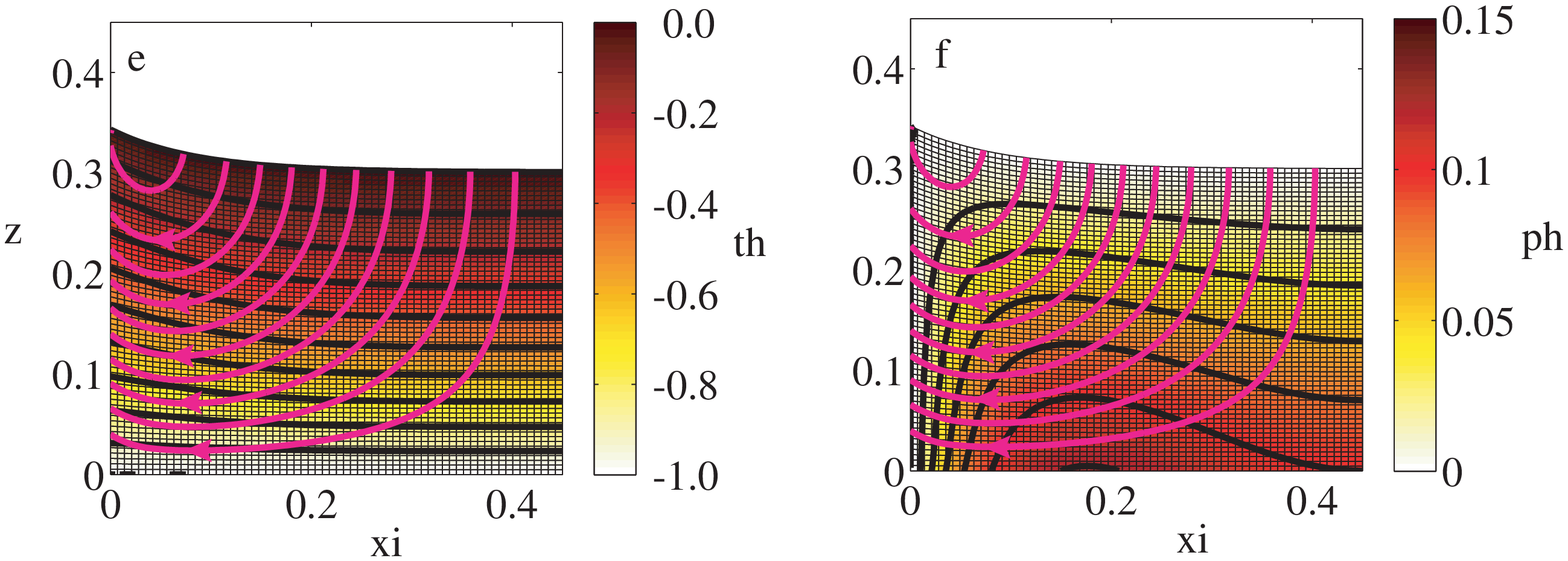} 
\caption{(Color online) Comparison of mushy region profiles at a long-wavelength chimney spacing $\lambda=1.0>\lambda_{\cal O}$ (\textit{a,b}), with those at the optimal chimney spacing $\lambda=\lambda_{\cal O}=0.46$ (\textit{c,d}). The temperature $\theta$ is indicated by the color scale in~(\textit{a,c}) with isotherms shown as solid black curves. Solid fraction $\phi$ is shown by the color scale in~(\textit{b,d}), constant $\phi$ contours as black curves, and Darcy velocity streamlines as magenta curves.
Other parameters are identical to fig.~\ref{fig:brinefluxes}.}
\label{fig:mushprofiles}
\end{figure}
Figs.~\ref{fig:mushprofiles}(\textit{a,b}) show profiles of steady state mushy layer temperature, solid fraction and streamlines of Darcy velocity for $\lambda=1.0$ at  $\lambda>\lambda_{\cal O}$.  At this large wavelength, approximately half of the mushy region is well drained by streamlines entering at the upper boundary and exiting through the chimney at $x=0$. However, there is a large nearly stagnant region away from the chimney, suggesting an explanation for the observed inefficient drainage  for large chimney spacings.  Compare this to the corresponding profiles for the optimal configuration at $\lambda=\lambda_{\cal O}$ (Figs.~\ref{fig:mushprofiles}\textit{c,d}) where the streamlines show efficient drainage via convective cells of order one aspect ratio. Thus, rather than drainage rate being controlled by buoyancy driven flow in the chimney, the optimal solute flux is controlled by the efficiency of convection {\em within} the mushy region. The temperature and solid fraction have qualitatively similar structure for both wavelengths, with significant horizontal variation of the latter leading to inhomogeneity in the concentration of the final material.

Having determined the system properties for a range of $\lambda$, we now apply the variational principle to select a preferred value of chimney spacing with maximal solute flux at $\lambda=\lambda_{\cal O}$ and calculate how the system varies with $\Rm$. The $\Rm\gg1$ simulations show that the optimal solute flux ${F_S}_{\cal O}$ increases approximately linearly with $\Rm$ suggesting the approximate scaling laws
\begin{align} F_S &=0 & \Rm&<\Rc \label{eq:BlowRmscale} \\
 F_S&\sim \gamma \left(\Rm-\Rc\right) & \Rm&\geq \Rc,  \label{eq:BhiRmscale}\end{align}
 for some constants $\gamma$ and $\Rc$ that depend on the other parameters imposed on the system. For $\Rm\gg1$ Eq. \eqref{eq:BhiRmscale} implies that the dimensional solute flux
  \begin{equation} \hat{F_S}\sim \gamma \rho_0 g  \beta (C_0-C_E)^2 \Pi_0/\mu, \qquad \Rm \gg 1, \label{eq:dimBflux} \end{equation}
 is independent of both the thermal diffusivity $\kappa$ and the solidification rate $V$. This is consistent with the rate of solute transport being controlled  by the large scale convective flow, independent of any effective transport induced by molecular diffusion. As a point of comparison, the heat flux in turbulent Rayleigh-B\'{e}nard convection is also predicted to be asymptotically independent of $\kappa$ in Kraichnan's ultimate strongly convective regime~\cite{KRAICHNAN:1962ad}.

The optimal chimney spacing $\lambda_{\cal O}$ and resulting mushy layer depth $h_{\cal O}$ both decrease as the Rayleigh number increases 
but the aspect ratio $\lambda_{\cal O}/h_{\cal O}$ asymptotes to a constant value for $\Rm \gg 1$. The stronger flow at larger $\Rm$  generates a thinner mushy layer, but the most efficient solute drainage is given by order one aspect ratio convective cells. This behavior is consistent with the constant mean aspect ratio observed in the transient phase of enthalpy method simulations~\cite{Katz:2008tg}. 

These results embolden us to suggest explanations for  phenomena observed during transient growth, such as growth from a fixed temperature boundary. Experiments show that, as the mushy layer thickens over time, extinction of convective flow in some of the chimneys leads to an increase of the mean spacing of chimneys \cite{Wettlaufer:1997rz}.  This coarsening may be consistent with the dynamics of optimal chimney spacing which we find has constant aspect ratio $\lambda_{\cal O}/h_{\cal O}$ for $\Rm \gg 1$.  This is consistent with the mean spacing of chimneys increasing with mushy layer depth during transient growth. Moreover, during transient growth, the extinction of flow in certain convective channels may be consistent with the flow stabilization found here for $\lambda \ll h$. Because the mean depth $h$ increases with fixed chimney spacing $\lambda$ during transient growth, the aspect ratio $\lambda/h$ decreases until it triggers a stabilization of convection and extinguishes flow in a selection of the chimneys. Taken together this offers a possible explanation for the observed mechanisms of the coarsening of chimney spacing as $h$ increases.  Comparison with previous work \cite{Notz:2009rm,Wettlaufer:1997rz} suggests that the scaling~\eqref{eq:BhiRmscale} may also be of relevance for transient growth at small concentration ratios $(\Cc \ll 1)$. In particular, consistent with experiments in a finite geometry~\cite[][]{Wettlaufer:1997rz}, the solute flux~\eqref{eq:dimBflux} predicts that the concentration of the liquid region will change approximately linearly in time. This would provide a simple parameterization of brine drainage from growing sea ice for use in large scale models without having to resolve natural horizontal variations in sea ice structure.

In summary, we have numerically analyzed strongly nonlinear convection in a solidifying mushy layer with a periodic array of chimneys with spacing  $\lambda$.  By varying $\lambda$, we have shown the existence of an optimal chimney spacing  $\lambda_{\cal O}$ that maximizes the solute flux from the mushy layer and hence also the rate of removal of its potential energy. This  $\lambda_{\cal O}$  yields convective cells of order one aspect ratio thereby efficiently draining the mushy layer, with weak flow for $\lambda\gg \lambda_{\cal O}$. For $\lambda\ll \lambda_{\cal O}$ there is stabilization, so that chimney convection cannot be supported for spacings smaller than a Rayleigh number dependent critical value, which suggests a method to suppress chimney formation in engineering applications. Steady states of chimney convection and no flow show hysteretic behavior. These mechanisms are consistent with dynamics controlled by a variational principle, with the spacing of chimneys adjusting to optimize the rate of release of potential energy from the mushy layer, and facilitate the most efficient route towards thermodynamic equilibrium. 
 
\begin{acknowledgments}

We thank the U.S. National Science Foundation Grant No. OPP0440841 and Yale University under the Bateman endowment for support of this research. 

\end{acknowledgments}

\bibliography{Wellsetalsubmitbibv2}

\begin{thebibliography}{10}%
\makeatletter
\providecommand \@ifxundefined [1]{%
 \ifx #1\undefined \expandafter \@firstoftwo
 \else \expandafter \@secondoftwo
\fi
}%
\providecommand \@ifnum [1]{%
 \ifnum #1\expandafter \@firstoftwo
 \else \expandafter \@secondoftwo
\fi
}%
\providecommand \enquote [1]{``#1''}%
\providecommand \bibnamefont  [1]{#1}%
\providecommand \bibfnamefont [1]{#1}%
\providecommand \citenamefont [1]{#1}%
\providecommand\href[0]{\@sanitize\@href}%
\providecommand\@href[1]{\endgroup\@@startlink{#1}\endgroup\@@href}%
\providecommand\@@href[1]{#1\@@endlink}%
\providecommand \@sanitize [0]{\begingroup\catcode`\&12\catcode`\#12\relax}%
\@ifxundefined \pdfoutput {\@firstoftwo}{%
 \@ifnum{\z@=\pdfoutput}{\@firstoftwo}{\@secondoftwo}%
}{%
 \providecommand\@@startlink[1]{\leavevmode\special{html:<a href="#1">}}%
 \providecommand\@@endlink[0]{\special{html:</a>}}%
}{%
 \providecommand\@@startlink[1]{%
  \leavevmode
  \pdfstartlink
   attr{/Border[0 0 1 ]/H/I/C[0 1 1]}%
   user{/Subtype/Link/A<</Type/Action/S/URI/URI(#1)>>}%
  \relax
 }%
 \providecommand\@@endlink[0]{\pdfendlink}%
}%
\providecommand \url  [0]{\begingroup\@sanitize \@url }%
\providecommand \@url [1]{\endgroup\@href {#1}{\urlprefix}}%
\providecommand \urlprefix [0]{URL }%
\providecommand \Eprint[0]{\href }%
\@ifxundefined \urlstyle {%
  \providecommand \doi [1]{doi:\discretionary{}{}{}#1}%
}{%
  \providecommand \doi [0]{doi:\discretionary{}{}{}\begingroup
  \urlstyle{rm}\Url }%
}%
\providecommand \doibase [0]{http://dx.doi.org/}%
\providecommand \Doi[1]{\href{\doibase#1}}%
\providecommand \bibAnnote [3]{%
  \BibitemShut{#1}%
  \begin{quotation}\noindent
    \textsc{Key:}\ #2\\\textsc{Annotation:}\ #3%
  \end{quotation}%
}%
\providecommand \bibAnnoteFile [2]{%
  \IfFileExists{#2}{\bibAnnote {#1} {#2} {\input{#2}}}{}%
}%
\providecommand \typeout [0]{\immediate \write \m@ne }%
\providecommand \selectlanguage [0]{\@gobble}%
\providecommand \bibinfo [0]{\@secondoftwo}%
\providecommand \bibfield [0]{\@secondoftwo}%
\providecommand \translation [1]{[#1]}%
\providecommand \BibitemOpen[0]{}%
\providecommand \bibitemStop [0]{}%
\providecommand \bibitemNoStop [0]{.\EOS\space}%
\providecommand \EOS [0]{\spacefactor3000\relax}%
\providecommand \BibitemShut [1]{\csname bibitem#1\endcsname}%
\bibitem{Wisdom}%
  \BibitemOpen
  \bibfield{author}{%
  \bibinfo {author} {\bibfnamefont{G.~J.}\ \bibnamefont{Sussman}}\ and\
  \bibinfo {author} {\bibfnamefont{J.}~\bibnamefont{Wisdom}},\ }%
  \emph{\bibinfo {title} {Structure and Interpretation of Classical
  Mechanics}}\ (\bibinfo {publisher} {MIT Press},\ \bibinfo {address} {Boston,
  MA},\ \bibinfo {year} {2001})%
  \bibAnnoteFile{NoStop}{Wisdom}%
\bibitem{Doering:2006tg}%
  \BibitemOpen
  \bibfield{author}{%
  \bibinfo {author} {\bibfnamefont{C.~R.}\ \bibnamefont{Doering}}, \bibinfo
  {author} {\bibfnamefont{F.}~\bibnamefont{Otto}},\ and\ \bibinfo {author}
  {\bibfnamefont{M.~G.}\ \bibnamefont{Reznikoff}},\ }%
  \bibfield{journal}{%
  \Doi{DOI 10.1017/S0022112006000097}{\bibinfo {journal} {J. Fluid Mech.}}\ }%
  \textbf{\bibinfo {volume} {560}},\ \bibinfo {pages} {229} (\bibinfo {year}
  {2006})%
  \bibAnnoteFile{NoStop}{Doering:2006tg}%
\bibitem{DOERING:1992vl}%
  \BibitemOpen
  \bibfield{author}{%
  \bibinfo {author} {\bibfnamefont{C.~R.}\ \bibnamefont{Doering}}\ and\
  \bibinfo {author} {\bibfnamefont{P.}~\bibnamefont{Constantin}},\ }%
  \bibfield{journal}{%
  \bibinfo {journal} {Phys. Rev. Lett.}\ }%
  \textbf{\bibinfo {volume} {69}},\ \bibinfo {pages} {1648} (\bibinfo {year}
  {1992})%
  \bibAnnoteFile{NoStop}{DOERING:1992vl}%
\bibitem{Worster:2000}%
  \BibitemOpen
  \bibfield{author}{%
  \bibinfo {author} {\bibfnamefont{M.~G.}\ \bibnamefont{Worster}},\ }%
  \emph{\bibinfo {title} {in Perspectives in fluid dynamics: a collective
  introduction to current research}}\ (\bibinfo {publisher} {Cambridge
  University Press},\ \bibinfo {address} {Cambridge},\ \bibinfo {year} {2000})\
  pp.\ \bibinfo {pages} {393--446}%
  \bibAnnoteFile{NoStop}{Worster:2000}%
\bibitem{Wettlaufer:1997rz}%
  \BibitemOpen
  \bibfield{author}{%
  \bibinfo {author} {\bibfnamefont{J.~S.}\ \bibnamefont{Wettlaufer}}, \bibinfo
  {author} {\bibfnamefont{M.~G.}\ \bibnamefont{Worster}},\ and\ \bibinfo
  {author} {\bibfnamefont{H.~E.}\ \bibnamefont{Huppert}},\ }%
  \bibfield{journal}{%
  \bibinfo {journal} {J. Fluid Mech.}\ }%
  \textbf{\bibinfo {volume} {344}},\ \bibinfo {pages} {291} (\bibinfo {year}
  {1997})%
  \bibAnnoteFile{NoStop}{Wettlaufer:1997rz}%
\bibitem{Worster:1997}%
  \BibitemOpen
  \bibfield{author}{%
  \bibinfo {author} {\bibfnamefont{M.~G.}\ \bibnamefont{Worster}},\ }%
  \bibfield{journal}{%
  \bibinfo {journal} {Annu. Rev. Fluid Mech.}\ }%
  \textbf{\bibinfo {volume} {29}},\ \bibinfo {pages} {91} (\bibinfo {year}
  {1997})%
  \bibAnnoteFile{NoStop}{Worster:1997}%
\bibitem{Peppinetal:2008}%
  \BibitemOpen
  \bibfield{author}{%
  \bibinfo {author} {\bibfnamefont{S.~S.~L.}\ \bibnamefont{Peppin}}, \bibinfo
  {author} {\bibfnamefont{H.~E.}\ \bibnamefont{Huppert}},\ and\ \bibinfo
  {author} {\bibfnamefont{M.~G.}\ \bibnamefont{Worster}},\ }%
  \bibfield{journal}{%
  \bibinfo {journal} {J. Fluid Mech.}\ }%
  \textbf{\bibinfo {volume} {599}},\ \bibinfo {pages} {465} (\bibinfo {year}
  {2008})%
  \bibAnnoteFile{NoStop}{Peppinetal:2008}%
\bibitem{Solomon:1998rc}%
  \BibitemOpen
  \bibfield{author}{%
  \bibinfo {author} {\bibfnamefont{T.~H.}\ \bibnamefont{Solomon}}\ and\
  \bibinfo {author} {\bibfnamefont{R.~R.}\ \bibnamefont{Hartley}},\ }%
  \bibfield{journal}{%
  \bibinfo {journal} {J. Fluid Mech.}\ }%
  \textbf{\bibinfo {volume} {358}},\ \bibinfo {pages} {87} (\bibinfo {year}
  {1998})%
  \bibAnnoteFile{NoStop}{Solomon:1998rc}%
\bibitem{Worster:1991}%
  \BibitemOpen
  \bibfield{author}{%
  \bibinfo {author} {\bibfnamefont{M.~G.}\ \bibnamefont{Worster}},\ }%
  \bibfield{journal}{%
  \bibinfo {journal} {J. Fluid Mech.}\ }%
  \textbf{\bibinfo {volume} {224}},\ \bibinfo {pages} {335} (\bibinfo {year}
  {1991})%
  \bibAnnoteFile{NoStop}{Worster:1991}%
\bibitem{SchulzeWorster:1998}%
  \BibitemOpen
  \bibfield{author}{%
  \bibinfo {author} {\bibfnamefont{T.~P.}\ \bibnamefont{Schulze}}\ and\
  \bibinfo {author} {\bibfnamefont{M.~G.}\ \bibnamefont{Worster}},\ }%
  \bibfield{journal}{%
  \bibinfo {journal} {J. Fluid Mech.}\ }%
  \textbf{\bibinfo {volume} {356}},\ \bibinfo {pages} {199} (\bibinfo {year}
  {1998})%
  \bibAnnoteFile{NoStop}{SchulzeWorster:1998}%
\bibitem{ChungWorster:2002}%
  \BibitemOpen
  \bibfield{author}{%
  \bibinfo {author} {\bibfnamefont{C.~A.}\ \bibnamefont{Chung}}\ and\ \bibinfo
  {author} {\bibfnamefont{M.~G.}\ \bibnamefont{Worster}},\ }%
  \bibfield{journal}{%
  \Doi{DOI 10.1017/S0022112001007558}{\bibinfo {journal} {J. Fluid Mech.}}\ }%
  \textbf{\bibinfo {volume} {455}},\ \bibinfo {pages} {387} (\bibinfo {year}
  {2002})%
  \bibAnnoteFile{NoStop}{ChungWorster:2002}%
\bibitem{EmmsFowler:1994}%
  \BibitemOpen
  \bibfield{author}{%
  \bibinfo {author} {\bibfnamefont{P.~W.}\ \bibnamefont{Emms}}\ and\ \bibinfo
  {author} {\bibfnamefont{A.~C.}\ \bibnamefont{Fowler}},\ }%
  \bibfield{journal}{%
  \bibinfo {journal} {J. Fluid Mech.}\ }%
  \textbf{\bibinfo {volume} {262}},\ \bibinfo {pages} {111} (\bibinfo {year}
  {1994})%
  \bibAnnoteFile{NoStop}{EmmsFowler:1994}%
\bibitem{SchulzeWorster:2005}%
  \BibitemOpen
  \bibfield{author}{%
  \bibinfo {author} {\bibfnamefont{T.~P.}\ \bibnamefont{Schulze}}\ and\
  \bibinfo {author} {\bibfnamefont{M.~G.}\ \bibnamefont{Worster}},\ }%
  \bibfield{journal}{%
  \bibinfo {journal} {J. Fluid Mech.}\ }%
  \textbf{\bibinfo {volume} {541}},\ \bibinfo {pages} {193} (\bibinfo {year}
  {2005})%
  \bibAnnoteFile{NoStop}{SchulzeWorster:2005}%
\bibitem{BriggsHensonMcCormick:2000}%
  \BibitemOpen
  \bibfield{author}{%
  \bibinfo {author} {\bibfnamefont{W.~L.}\ \bibnamefont{Briggs}}, \bibinfo
  {author} {\bibfnamefont{V.~E.}\ \bibnamefont{Henson}},\ and\ \bibinfo
  {author} {\bibfnamefont{S.~F.}\ \bibnamefont{McCormick}},\ }%
  \emph{\bibinfo {title} {A Multigrid Tutorial}}\ (\bibinfo {publisher}
  {SIAM},\ \bibinfo {year} {2000})%
  \bibAnnoteFile{NoStop}{BriggsHensonMcCormick:2000}%
\bibitem{Adams:1989}%
  \BibitemOpen
  \bibfield{author}{%
  \bibinfo {author} {\bibfnamefont{J.~C.}\ \bibnamefont{Adams}},\ }%
  \bibfield{journal}{%
  \bibinfo {journal} {Appl. Math. Comput.}\ }%
  \textbf{\bibinfo {volume} {34}},\ \bibinfo {pages} {113} (\bibinfo {year}
  {1989})%
  \bibAnnoteFile{NoStop}{Adams:1989}%
\bibitem{Keller:1977}%
  \BibitemOpen
  \bibfield{author}{%
  \bibinfo {author} {\bibfnamefont{H.~B.}\ \bibnamefont{Keller}},\ }%
  \emph{\bibinfo {title} {in Applications of Bifurcation Theory}}\ (\bibinfo
  {publisher} {Academic Press},\ \bibinfo {address} {New York},\ \bibinfo
  {year} {1977})\ pp.\ \bibinfo {pages} {359--384}%
  \bibAnnoteFile{NoStop}{Keller:1977}%
\bibitem{KRAICHNAN:1962ad}%
  \BibitemOpen
  \bibfield{author}{%
  \bibinfo {author} {\bibfnamefont{R.~H.}\ \bibnamefont{Kraichnan}},\ }%
  \bibfield{journal}{%
  \bibinfo {journal} {Phys. Fluids}\ }%
  \textbf{\bibinfo {volume} {5}},\ \bibinfo {pages} {1374} (\bibinfo {year}
  {1962})%
  \bibAnnoteFile{NoStop}{KRAICHNAN:1962ad}%
\bibitem{Katz:2008tg}%
  \BibitemOpen
  \bibfield{author}{%
  \bibinfo {author} {\bibfnamefont{R.~F.}\ \bibnamefont{Katz}}\ and\ \bibinfo
  {author} {\bibfnamefont{M.~G.}\ \bibnamefont{Worster}},\ }%
  \bibfield{journal}{%
  \Doi{DOI 10.1016/j.jcp.2008.06.039}{\bibinfo {journal} {J. Comput. Phys.}}\
  }%
  \textbf{\bibinfo {volume} {227}},\ \bibinfo {pages} {9823} (\bibinfo {year}
  {2008})%
  \bibAnnoteFile{NoStop}{Katz:2008tg}%
\bibitem{Notz:2009rm}%
  \BibitemOpen
  \bibfield{author}{%
  \bibinfo {author} {\bibfnamefont{D.}~\bibnamefont{Notz}}\ and\ \bibinfo
  {author} {\bibfnamefont{M.~G.}\ \bibnamefont{Worster}},\ }%
  \bibfield{journal}{%
  \bibinfo {journal} {J. Geophys. Res.-Oceans}\ }%
  \textbf{\bibinfo {volume} {114}},\ \bibinfo {pages} {C05006} (\bibinfo {year}
  {2009})%
  \bibAnnoteFile{NoStop}{Notz:2009rm}%
\end{thebibliography}%

\end{document}